# Three regimes of extrasolar planets inferred from host star metallicities

Lars A. Buchhave[1,2], Martin Bizzarro[2], David W. Latham[1], Dimitar Sasselov[1], William D. Cochran[3], Michael Endl[3], Howard Isaacson[4], Diana Juncher[2,5] & Geoffrey W. Marcy[4]

[1]Harvard-Smithsonian Center for Astrophysics, Cambridge, Massachusetts 02138, USA.

[2]Centre for Star and Planet Formation, Natural History Museum of Denmark, University of Copenhagen, DK-1350 Copenhagen, Denmark.

[3]McDonald Observatory, The University of Texas, Austin, Texas 78712, USA.

[4]University of California, Berkeley, California 94720, USA.

[5]Niels Bohr Institute, University of Copenhagen, DK-2100 Copenhagen, Denmark.

**Approximately half of the extrasolar planets (exoplanets) with radii less than four Earth radii are in orbits with short periods[1]. Despite their sheer abundance, the compositions of such planets are largely unknown. The available evidence suggests that they range in composition from small, high-density rocky planets to low-density planets consisting of rocky cores surrounded by thick hydrogen and helium gas envelopes. Understanding the transition from the gaseous planets to Earth-like rocky worlds is important to estimate the number of potentially habitable planets in our Galaxy and provide constraints on planet formation theories. Here we report the abundances of heavy elements (that is, the metallicities) of more than 400 stars hosting 600 exoplanet candidates, and find that the exoplanets can be categorized into three populations defined by statistically distinct (~ 4.5σ) metallicity regions. We interpret these regions as reflecting the formation regimes of terrestrial-like planets (radii less than 1.7 Earth radii), gas dwarf planets with rocky cores and hydrogen–helium envelopes (radii between 1.7 and 3.9 Earth radii) and ice or gas giant planets (radii greater than 3.9 Earth radii). These transitions correspond well with those inferred from dynamical mass estimates[2,3], implying that host star metallicity, which is a proxy for the initial solid inventory of the protoplanetary disk, is a key ingredient regulating the structure of planetary systems.**

Shortly after the discovery of the first exoplanets, host star metallicity was suggested to have a role in the formation of planetary systems[4]. Indeed, the well-established tendency for hot Jupiters to be found more frequently orbiting metal rich stars has been confirmed by a number of studies[5,6]. Although it has recently been shown that small planets form for a wide range of host star metallicities[7–11], it is clear that the metallicities of stars with small planets are on average lower than those of gas giants. This suggests that subtle differences may exist in the metallicities of the host stars of small exoplanets, and this, in turn, may be linked to distinct physical properties of the underlying planet populations. However, effectively probing this regime requires a large sample of homogeneously derived metallicities for stars with small planets. Therefore, using our stellar parameters classification (SPC) tool[7] (Methods Summary), we analyse more than 2,000 high-resolution spectra of Kepler Objects of Interest[12] (KOIs) gathered by the Kepler Follow-up Program, yielding precise stellar parameters (provided as a table in machine-readable form), including metallicities, of 405 stars orbited by 600 exoplanet candidates.

Our sample of spectroscopic metallicities of stars hosting small planets is a factor of two larger than any previous sample[7], allowing us to probe in greater detail for significant differences in the metallicities of stars hosting planets of different sizes. At various radii, we divide the sample into two bins of stars hosting small and large planets, and perform a two-sample Kolmogorov–Smirnov test to determine whether the metallicities of the two distributions of host stars are not drawn randomly from the same parent population. We find two significant features in the Kolmogorov–Smirnov test diagram, one at $1.7 R_\oplus$ with a significance of $4.5\sigma$ and one at $3.9 R_\oplus$ with a significance of $4.6\sigma$, suggesting transitions between three exoplanet size regimes (Fig. 1). The average metallicity of the host stars increases with planet size, yielding average metallicities of $-0.02 \pm 0.02$, $0.05 \pm 0.01$ and $0.18 \pm 0.02$ dex in the respective regimes. To assess the uncertainty in radius at which these transitions occur, we perform a Monte Carlo simulation by drawing $10^6$ sets of data, where the host star metallicities and planetary radii are randomly perturbed within the uncertainties (the uncertainty in the planetary radius is assumed to be dominated by the uncertainty in the radius of the host star). We find the two features to be at $1.55^{+0.88}_{-0.04} R_\oplus$ ($4.2^{+0.5}_{-0.4}\sigma$) and $3.52^{+0.74}_{-0.28} R_\oplus$ ($4.7^{+0.6}_{-0.4}\sigma$), consistent with the original data.

Small planets with short periods could undergo significant evaporation of their atmospheres[13]. These planets will therefore not obey the radius–metallicity relation we are studying, because any accumulated gas would have evaporated. Therefore, we remove small ($R_P < 3 R_\oplus$, where $R_P$ is the exoplanet radius), highly irradiated planets (stellar flux, $F_v > 5 \times 10^5$ J s$^{-1}$ m$^{-2}$) from the sample, leaving 463 planets orbiting 324 stars, which increases the significance of the feature at $1.7R_\oplus$ from $3.5\sigma$ to the reported $4.5\sigma$. For comparison, the rocky planet Kepler-10b, with an 0.8-d period[14], receives a flux of $F_v \approx 48 \times 10^5$ J s$^{-1}$ m$^{-2}$, whereas Kepler-11c[15], whose density suggests it is gaseous, receives $F_v \approx 1.3 \times 10^5$ J s$^{-1}$ m$^{-2}$.

Recent studies suggest that the masses and radii of small planets ($1.5R_\oplus$–$4R_\oplus$) follow a linear relationship, implying that planet density decreases with increasing planet radius[2]. However, this relationship must change significantly for larger planets (>$4R_\oplus$) to explain the large mass of gas giant planets such as Jupiter[2]. Our data indicate a statistically significant increase in metallicity at a comparable planetary radius, $R_P = 3.9R_\oplus$. This observation is in agreement with the well-established correlation between a star's metallicity and its likelihood to host hot Jupiters[5,6], confirming that the formation regime for larger planets (>$3.9R_\oplus$) requires exceptionally high metallicity environments[7]. We therefore interpret the regime of larger planets ($R_P > 3.9R_\oplus$) to consist of ice and gas giant planets formed beyond the 'snow line' at around ~3 AU (1 AU is the average Sun–Earth distance), where the availability of solids is a factor of four higher because volatile elements are able to condense and form solids at cooler temperatures[16]. The high concentration of heavy elements in the protoplanetary disk, resulting from the higher metallicity and the condensation of volatiles, and the planet's large distance to the host star allow these planets to grow rapidly and amass a gaseous atmosphere before the gas in the protoplanetary disk dissipates. These planets then migrate to their present positions much closer to their host stars[17], yielding the hot Jupiters seen orbiting stars with high metallicities.

To explore the implications of the feature at $1.7R_\oplus$ in the metallicity–radius plane (Fig. 1), we compare our results with recent work attempting to determine the radius at which the transition from gaseous to rocky planets occurs. Dynamical masses derived from precise radial velocities of transiting exoplanets indicate that planets with $R_P > 2R_\oplus$ have densities that imply increasing amounts by volume of light material, whereas planets with $R_P < 1.5R_\oplus$ have densities systematically greater than that of Earth[2]. Moreover, an analysis of data for a larger sample of planets (including masses derived from transit timing variations), has shown that planets with $R_P < 1.5R_\oplus$ probably are of rocky composition[3]. Finally, it has been suggested that $R_P \approx 1.75R_\oplus$ is a physically motivated transition point between rocky and gaseous planets, based on reported masses and radii combined with thermal evolutionary atmosphere models[18]. The statistically significant peak in the metallicity–radius plane at $1.7R_\oplus$ agrees with these findings, suggesting that the compositions of small exoplanets ($R_P < 3.9R_\oplus$) in close proximity to their host stars are also regulated by the number density of solids in the protoplanetary disk. Thus, we interpret the two regimes of smaller planets identified by the host star metallicities as reflecting the transition between rocky terrestrial exoplanets that have not amassed a gaseous atmosphere ($R_P < 1.7R_\oplus$) and planets with rocky cores that have accumulated an envelope of hydrogen, helium and other volatiles, which we denote gas dwarfs ($1.7R_\oplus < R_P < 3.9R_\oplus$).

The formation mechanism of the terrestrial and gas dwarf exoplanet regimes in short orbital periods is not fully understood. In one model, these small exoplanets are believed to form *in situ* with little post-assembly migration[19,20]. Although the *in situ* accretion model seems to be successful in reproducing the observed distribution of the 'hot Neptune' and super-systems, including their orbital spacing[21], it requires unusually large amounts of solids in the innermost protoplanetary disk. A competing model invokes accretion during the inward migration of a population of planetary embryos formed at a range of orbital distances beyond the ice line[22,23]. On this view, Mars- and Earth-size embryos migrate inwards owing to tidal interaction with the disk[24], and accumulate at the inner edge of the protoplanetary disk, where they complete their assembly[25]. Irrespective of the formation mechanism, however, the observed peak in the metallicity–radius plane at $1.7R_\oplus$ suggests that the final mass and composition of a small exoplanet is controlled by the amount of solid material available in the protoplanetary disk. A higher-metallicity environment promotes a more rapid and effective accretion process, thereby allowing the cores to amass a gaseous envelope before dissipation of the gas. In contrast, lower-metallicity environments may result in the assembly of rocky cores of several Earth masses on timescales greater than that inferred for gas dispersal in protoplanetary disks[19] (<10 Myr), yielding cores without gaseous hydrogen–helium atmospheres.

A prediction from gas accretion on short orbital periods is that the critical mass at which a core can accrete an atmosphere is $M_{cr} \approx 2.6 M_\oplus (\eta/0.3)^{1/2} (P_{orb}/1\,d)^{5/12}$, where $\eta = M_{atm}/M_{cr}$ is the fractional mass comprised by the atmosphere[26]. In this model, the planetary mass and, thus, radius indicating the transition from rocky to gaseous planets should increase with orbital period. However, the exact opposite dependence, namely a decrease in core mass with increased orbital period, has also been suggested[27]. To investigate whether the radius of transition from rocky to gaseous planets found in our data shows a dependence on orbital period, we segregate the sample by period into four bins with approximately equal numbers of planets. We repeat the previously described Kolmogorov–Smirnov test for the planets in each of the period bins: we remove the larger planets ($R_P > 3.9 R_\oplus$) and perform a Monte Carlo simulation by drawing $10^6$ sets of data, where the host star metallicities and planetary radii are randomly perturbed within the uncertainties. The red line in Fig. 2 is a power-law fit to the transition radius, $R_{cr}$, inferred from our data ($R_{cr} = 1.06 P_{orb}^{0.17}$ Rcr = 1.06 $R_\oplus$ ($P_{orb}$/1 day)$^{0.17}$). Although additional data are required to confirm this relationship, the fit is apparently consistent with a critical core mass that increases with orbital period and an atmospheric fraction of 5% (ref. 26; blue dashed line in Fig. 2). If correct, this predicts the existence of more massive rocky exoplanets at longer orbital periods.

Although our analysis of a statistically significant number of planets and their host star metallicities allows us to distinguish between three distinct exoplanet regimes, we emphasize that a multitude of factors can affect the outcome of planet formation. Therefore, the transition radii inferred from our analysis probably represent gradual transitions between the different planet regimes and so may not apply to all planetary systems. However, the agreement between the transition radii inferred here and those deduced from dynamical mass measurements of transiting planets[2,3] implies that host star metallicity—and, by extension, the solid inventory of a protoplanetary disk—is one of the driving factors determining the outcome of planet formation.

**METHODS SUMMARY**

We use SPC[7] for the spectroscopic analysis yielding the stellar parameters in this work. SPC uses a grid of synthetic library spectra to derive the effective temperature, surface gravity, metallicity and rotational velocity simultaneously by matching the models with the observed spectra originating from a number of different instruments (Methods). The stellar parameters from SPC and Yonsei–Yale stellar evolutionary models[28] are used to estimate the radii of the host stars, which we couple with the photometric data from the Kepler mission to infer the planet radii. The majority of planets in the sample have short orbital periods, owing to the observational bias of Kepler towards shorter period planets. The mean and median periods are 38.0 and 12.4 d, respectively.

To investigate whether the radius of transition from rocky to gaseous planets found in our data shows a dependence on orbital period (Fig. 2), we remove the larger planets ($R_P > 3.9R_\oplus$) and segregate the remainder of the sample by period into four bins with approximately equal numbers of planets. The previously described Kolmogorov–Smirnov test (Fig. 1) is repeated for each bin by performing a Monte Carlo simulation in which the host star metallicities and planetary radii are randomly perturbed within the uncertainties. The data points plotted in Fig. 2 are the means of the resulting posterior distributions, and the error bars indicate the $1\sigma$ uncertainties.

To ensure that the statistical analyses are sound, we perform further tests evaluating the effects of uneven sampling and contamination (Methods). We find results consistent with those presented in the paper within the uncertainties, and conclude that our statistical analyses are robust.

**Supplementary Information** is available in the online version of the paper.

**Acknowledgements** L.A.B. acknowledges support from the Harvard Origins of Life Initiative. M.B. acknowledges funding from the Danish National Research Foundation (grant number DNRF97) and from the European Research Council under ERC Consolidator grant agreement 616027- STARDUST2ASTEROIDS. D.W.L. acknowledges support from the Kepler Mission under NASA Cooperative Agreements NCC2-1390, NNX11AB99A and NNX13AB58A with the Smithsonian Astrophysical Observatory, and thanks the observers who helped obtain the TRES observations reported here, especially R. Stefanik, G. Esquerdo, P. Berlind and M. Calkins.

**Author Contributions** L.A.B. led the project and developed the classification tools for the metallicity analysis. M.B., D.W.L. and D.S. contributed to the discussion of the theoretical implications of the data. L.A.B., D.W.L., W.D.C., M.E., H.I., D.J. and G.W.M. worked on gathering the spectroscopic observations. All authors discussed the results and commented on the manuscript. L.A.B. and M.B. wrote the paper with input from D.W.L. and D.S.

**Author Information** Reprints and permissions information is available at www.nature.com/reprints. The authors declare no competing financial interests. Readers are welcome to comment on the online version of the paper. Correspondence and requests for materials should be addressed to L.A.B. (lbuchhave@cfa.harvard.edu).

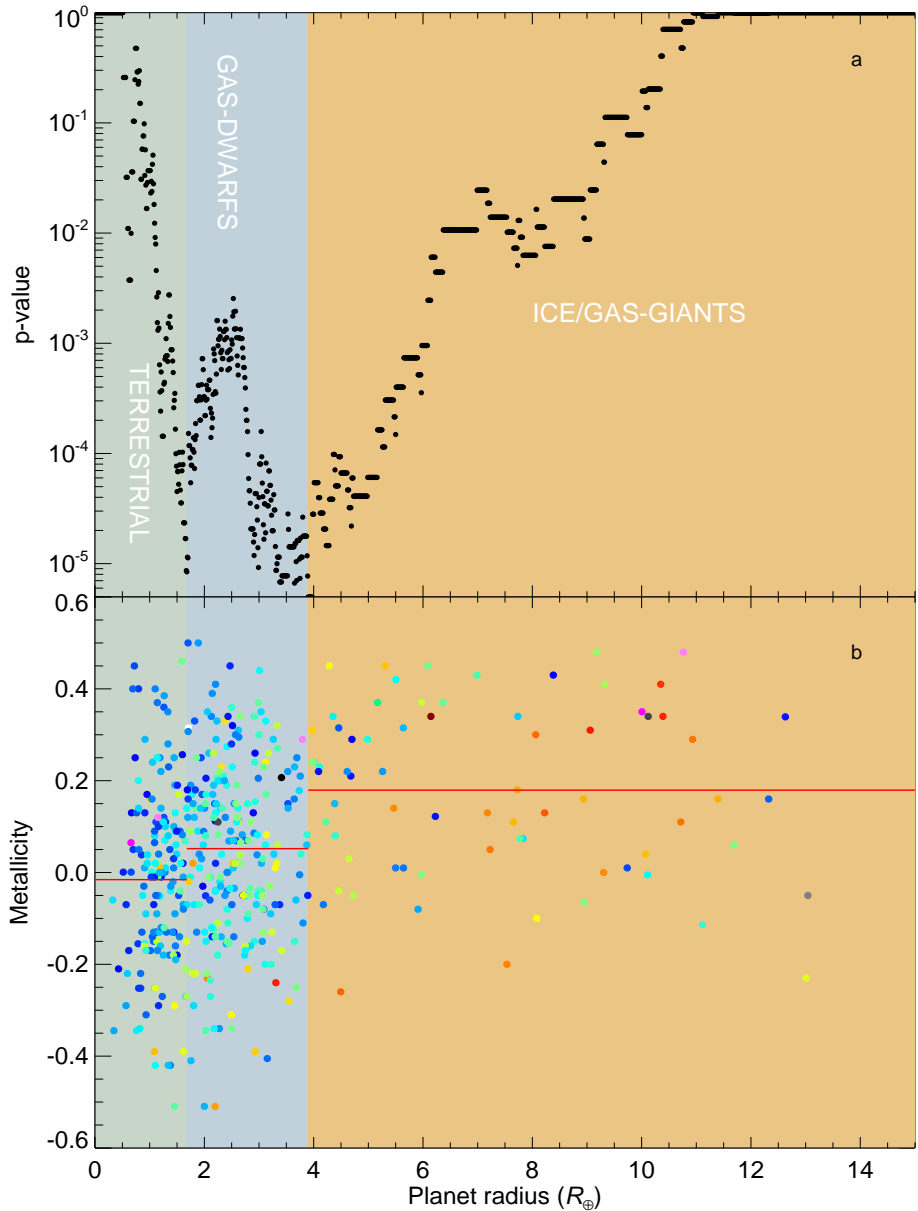

**Figure 1 Host star metallicities and three types of exoplanets with different composition. a**, *P* value of the two-sample Kolmogorov−Smirnov test. **b**, Radii of the individual planets and their host star metallicities. Point colour represents the logarithm of the period of the planets (blue, shortest period; red, longest period). The solid red lines are the average metallicities in the three regions (−0.02 ± 0.02, 0.05 ± 0.01 and 0.18 ± 0.02 dex, where each uncertainty is 1 s.e.m. of the host star metallicities in the corresponding bin).

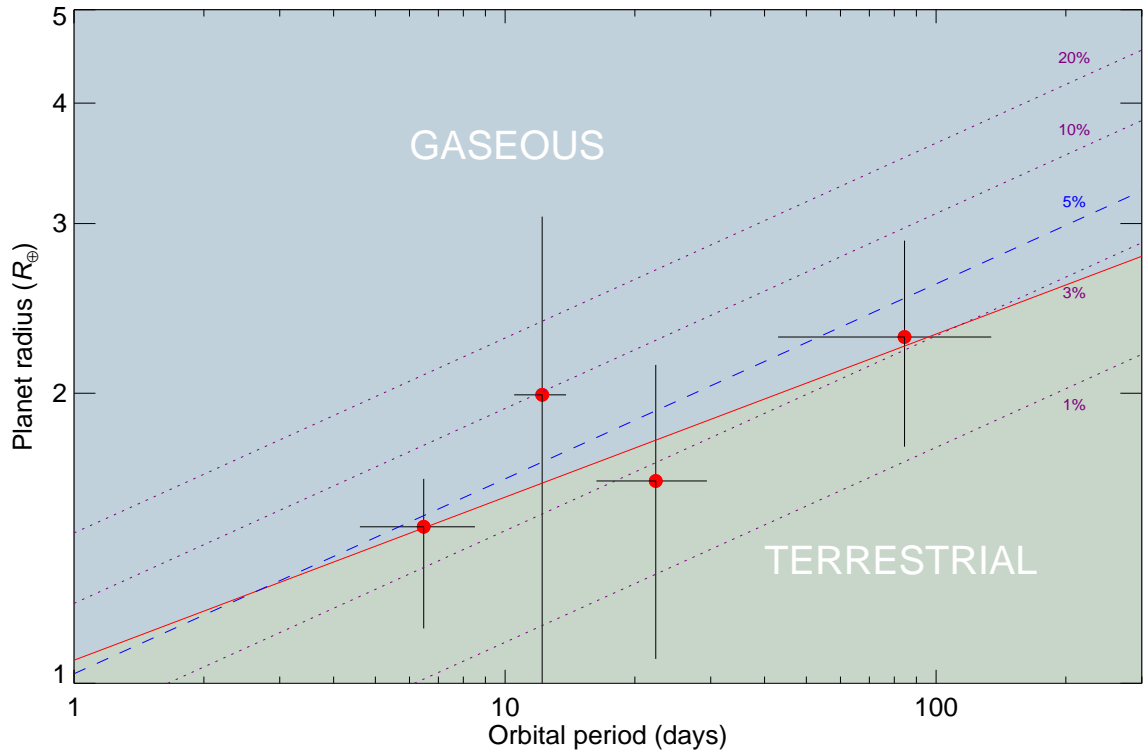

**Figure 2 The radius of transition from rocky to gaseous exoplanets.** The transition radii (red points) are the means of the posterior distributions resulting from the Monte Carlo analysis (main text), and the error bars indicate the $1\sigma$ uncertainties. Using the mass–radius approximation[29] $M/M_\oplus = (R/R_\oplus)^{2.06}$, we plot the radius corresponding to $M_{cr}$ with an atmosphere fraction of 5% as the blue dashed line and those for respective atmosphere fractions of 1%, 3%, 10% and 20% as the dotted purple lines. The solid red line is a power-law fit to the Monte Carlo data: $R_{cr} = 1.06 P_{orb}^{0.17}$  Rcr = 1.06 R$_\oplus$ (P$_{orb}$/1 day)$^{0.17}$.

## METHODS

**Observations and stellar parameters**

This study is based on stellar classifications by SPC[7] of 2,297 spectra observed using the Fibre-fed Echelle Spectrograph on the 2.6-m Nordic Optical Telescope on La Palma, Spain (488 spectra), the fibre-fed Tillinghast Reflector Echelle Spectrograph on the 1.5-m Tillinghast Reflector at the Fred Lawrence Whipple Observatory on Mt Hopkins, Arizona (985 spectra), the Tull Coudé Spectrograph on the 2.7-m Harlan J. Smith Telescope at the McDonald Observatory Texas (653 spectra) and the HIRES spectrograph on the 10-m Keck I telescope at Mauna Kea, Hawaii (171 spectra). We included only the most secure stellar classifications by limiting our sample to stars with effective temperatures of 4,800 K < $T_{eff}$ < 6,500 K, projected rotational velocities of $v\sin(i) < 20$ km s$^{-1}$, spectra with signal-to-noise ratios per resolution element of more than 25, and normalized cross-correlation function peak heights of more than 0.9 (indicating the quality of the stellar classification).

We improve on the determination of the surface gravity, known to be prone to degeneracies with effective temperature and metallicity[30], by imposing a prior on the surface gravity from stellar evolutionary models and an initial estimate of the star's effective temperature and metallicity. This is particularly useful for cooler stars, where the evolutionary models put tight constraints on the surface gravity. We use the stellar parameters from SPC and the Yonsei–Yale stellar evolutionary models[28] to estimate the radii of the host stars, and, using the photometrically derived planet radii from Kepler, we correct the planetary radii based on the Kepler Input Catalogue photometry, which are known to be prone to systematic biases and large uncertainties. The improved stellar radii reduce the uncertainties in the planetary radii from an average error of 34% to one of 11%, assuming that the major contribution to the uncertainty in the planetary radii originates from the stellar radii.

**Uneven sampling**

To investigate the effect of uneven sample size, we performed a Monte Carlo test with $10^6$ realizations where we randomly drew observations from the smaller of the two samples, making each of the two samples equal in size at all times. We find the ice or gas giant transition to be at $3.58^{+0.75}_{-0.38} R_\oplus$ with a significance of $4.9^{+0.5}_{-0.4}\sigma$ and the rocky transition to be at $1.60^{+0.83}_{-0.10} R_\oplus$ with a significance of $4.4^{+0.5}_{-0.4}\sigma$. Both values are consistent with the Monte Carlo analysis reported in the paper. We conclude that the uneven sample size does not affect the significance of our statistical analysis.

**Contamination**

We are using the two-sample Kolmogorov–Smirnov test for the statistical analysis, but we find three distinct populations of exoplanets. To establish whether contamination from the third sample affects our results, we remove the planets from the third sample (removing $R_P > 3.9 R_\oplus$ when searching for the peak at $1.7 R_\oplus$ and removing $R_P < 1.7 R_\oplus$ for the peak at $3.9 R_\oplus$) and subsequently carry out the Kolmogorov–Smirnov test in an attempt to recover each of the peaks in Fig. 1. We find the first transition to be at the same radius ($1.7 R_\oplus$), albeit with a lower significance ($3.1\sigma$). We find the ice/gas giant transition close to the one reported in the paper, again with a slightly lower significance ($4.2 R_\oplus$ at $3.9\sigma$). Again, evaluation of our data using a different approach supports our results and conclusions.